\documentclass[final, a4paper]{aastex631}

\usepackage{color}
\usepackage{enumitem}
\usepackage{hyperref}
\usepackage{verbatim}
\usepackage{amsmath,amstext,mathrsfs}
\usepackage[all]{hypcap} 
\usepackage{afterpage}    
\usepackage{marginnote}
\usepackage{makecell}
\usepackage{subfigure}
\usepackage{multirow}

\shorttitle{Energy  Equipartition}
    \shortauthors{Li \& Zhao}
\usepackage{enumitem}
\setlist[enumerate]{listparindent=\parindent}
\makeatletter

\newcommand{\Rmnum}[1]{\expandafter\slowromancap\romannumeral #1@}
\makeatother

\begin{document}
    \hfuzz = 150pt
\title{\Large\bfseries Volume Density Mapper: 3D Density Reconstruction Algorithm for Molecular Clouds }

\author[0000-0003-3144-1952]{Guang-Xing Li}
\affil{South-Western Institute for Astronomy Research, Yunnan University, Kunming 650091, People’s Republic of China}

\author[0000-0003-0596-6608]{Mengke Zhao}
\affil{School of Astronomy and Space Science, Nanjing University, 163 Xianlin Avenue, Nanjing 210023, Jiangsu, People’s Republic of China}
\affil{Key Laboratory of Modern Astronomy and Astrophysics (Nanjing University), Ministry of Education, Nanjing 210023, Jiangsu, People’s Republic of China}

\begin{abstract}
The interstellar medium (ISM) exhibits complex, multi-scale structures that are challenging to study due to their projection into two-dimensional (2D) column density maps. We present the Volume Density Mapper, a novel algorithm based on constrained diffusion to reconstruct three-dimensional (3D) density distributions of molecular clouds from 2D observations. This method decomposes the column density into multi-scale components, reconstructing a 3D density field that preserves key physical properties such as mean density, maximum density, and standard deviation along the line of sight. Validated against numerical simulations (FLASH and ENZO), the algorithm achieves high accuracy, with mean density estimates within 0.1 dex and dispersions of 0.2 to 0.3 dex across varied cloud structures. The reconstructed 3D density fields enable the derivation of critical parameters, including volume density, cloud thickness, and density probability distribution functions, offering insights into star formation and ISM evolution. The versatility of the method is demonstrated by applying diverse systems from galaxies (NGC 628) to protostellar disks. The code is available at https://github.com/gxli/volume-density-mapper. 
\end{abstract}

\section{Introduction}

The interstellar medium (ISM)  is a complex system where the gas clouds exhibit
complex self-similar hierarchies of structures. Understanding the evolution of
these complex systems remain challenging due to the complexity, complicated by
the fact that these systems are often observed in projection, where only 2D
surface density distribution is available.

Estimating the 3D density distribution of ISM clouds is crucial for
understanding the evolution, for several reason. The density field offers
a complete description of the matter distribution, and understanding of the 3D
density distribution can be used to gain understanding of 
such as pressure balance and gravitational collapse, and tidal force \cite{2014Sci...344..183K}. In the past, methods have
been proposed to estimate the 3D density distribution from the 2D surface
density distribution.  Models such as the  spherical models
\citep{2020A&A...633A.132H} do offer such possibility, however, the overly
simple assumption hinders widespread adoption. Using the wavelet transform to
capture the multi-scale nature of clouds. However, the wavelet transform
performs poorly for high-contract images, such as the clouds which often have
log-normal or power-law like density distributions. Recently methods \citep{2025arXiv250319259L}  can take
the filamentary nature of the clouds into account. However, the computations can
be expensive. 

We introduce the constrained diffusion 
\citep{2022ApJS..259...59L}-based approach for reconstructing 3D
density fields from column density observations. We note that due to the
projection effect, it is impossible to completely recover the 3D density
distribution. The reconstruction does not at attempt to  recover the full
3D density. Rather, it attempt to recover the 3D density distribution that
resemble the true 3D density distribution, in that key quantifies such as the
mean density, maximum density, and standard deviation of the density
distribution are kept consistent. This is achieved using the constrained
diffusion  method  \citep{2022ApJS..259...59L} in a direct, transparent fashion.
In this paper we describe the method and its performance, and demonstrate its
applicability to a wide range of ISM clouds. A implementation of the method is
available at \url{https://github.com/gxli/volume-density-mapper}.


\section{Method}

Clouds are complex structures that span multiple scales. The reconstruction
represent here is a reconstruction which aims to obtain a 3D density
distribution that have the following properties: (1) the 3D density
distribution, when projected on 2D
is consistent with the surface density; (2) alone each line of sight, the 3D
density distribution is smooth and continuous; (3) alone each line of sight, the
3D density distribution resembles that of the 2D density distribution, where
maximum density, mean density and standard deviation are consistent with the
real density distribution. 

The method consists of two steps: (1) decompose the surface density using the constrained diffusion algorithm, and (2) reconstruct the 3D density distribution from the decomposed components. 
\subsection{Constrained Diffusion: A multi-scale representation of cloud structures}

 To illustrate the principle of the 3D density distribution, we use the equation,
\begin{equation}
\rho(x,y,z) = \Sigma_i (x,y) / l_i
\end{equation}
to estimate the 3D density distribution, where $ \Sigma_i (x,y)$ is the decomposed component from original column density map and the $l_i$ is the specific scale of the decomposed component.
The major difficulty in estimating the scale is the cloud.  Unlike a slab or a
sphere, clouds have complex density structures. Therefore, to convert the 2D surface
density distribution to a 3D density distribution, it is impossible to define
a unique scale. To address this challenge, we use the method of constrained
diffusion \citep{2022ApJS..259...59L} to decompose the surface density distribution into components at
different scales, and obtained the 3D density distribution by using these
component maps at different scales. This approach has been used in
\citet{2025arXiv250801130Z} to construct volume density distribution maps.

To construct the 3D density distribution, we first decompose the surface
density.

Toward each scale, the
surface density distribution is
\begin{equation}
    \Sigma_l(x,y)\;,\nonumber
\end{equation}
where the density can be estimated as
\begin{equation}
    \rho_l(x,y,z) = \Sigma_l(x,y) H(l,z)\;,
\end{equation}
where $H(l,z)$ is a kernel function that describes the height of the region at
scale $l$. Since the constrained diffusion algorithm is designed to separate
different Gaussian-like structures, we choose our kernel function to be
\begin{equation}\label{eq:thick}
    H(l,z) =  \frac{1}{\sqrt{2\pi}l  f_z} \exp\left(-\frac{z^2}{2(lf_z)^2}\right)\;,
\end{equation}
where $f_z =1$ is a normalization factor determined empirically through comparing
our result with numerical simulations.

The implementation of method is rather straightforward. It consists of the
following steps: 
\begin{itemize}
    \item Decompose the surface density map into components at different scales using
    the constrained diffusion algorithm \citep{2022ApJS..259...59L}.
    \item For each component, estimate produce a 3D density distribution using the
    equation \ref{eq:thick}, and store the 3D density distribution in a 3D cube.
    \item Assuming the 3D density cubes from the previous cubes, producing the
    final density reconstruction.
\end{itemize} 

A version of the implementation is available at
\url{https://github.com/gxli/volume-density-mapper}.
Our method estimates the size of the region using its
shape observed in projection. Thus, accurate at a given location is depended on
the underlying 3D shape. When the density structure is singular, e.g. a
completely straight filament is viewed from its end, the reconstruction can have
significant errors.  


\begin{figure}
    \centering
    \includegraphics[width = \textwidth]{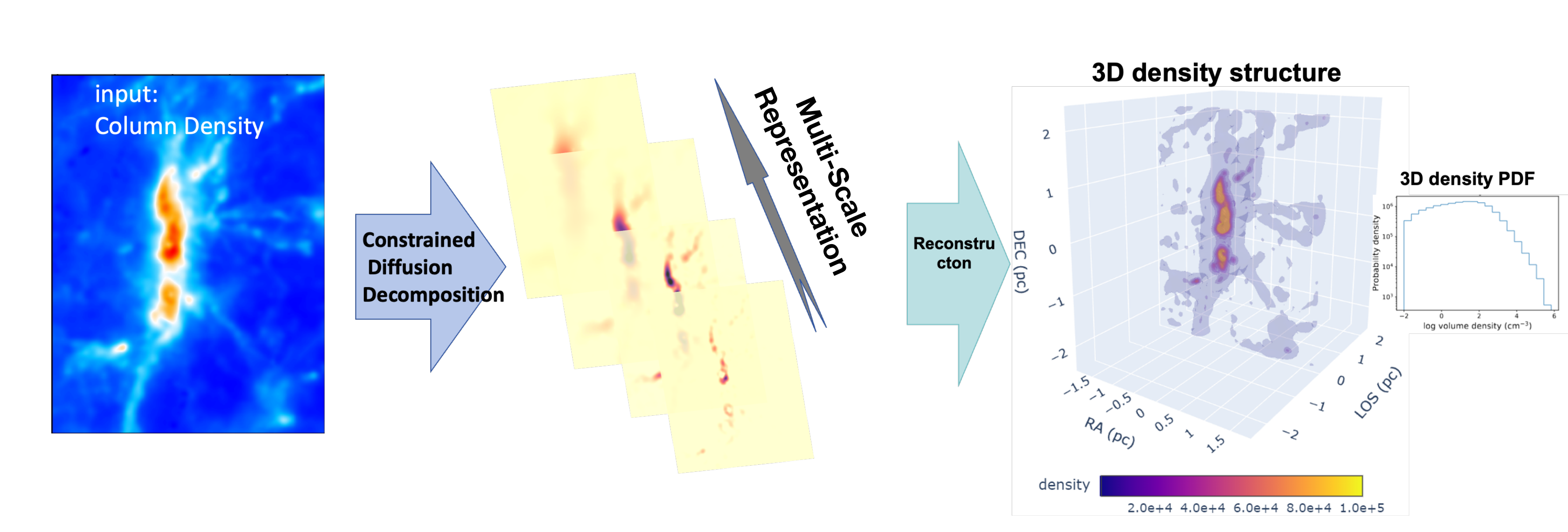}
    \caption{ \bf Flowchart of the density reconstruction method.}
    \label{fig:flowchart}
\end{figure}

\begin{figure}
    \centering
    \includegraphics[width=\linewidth]{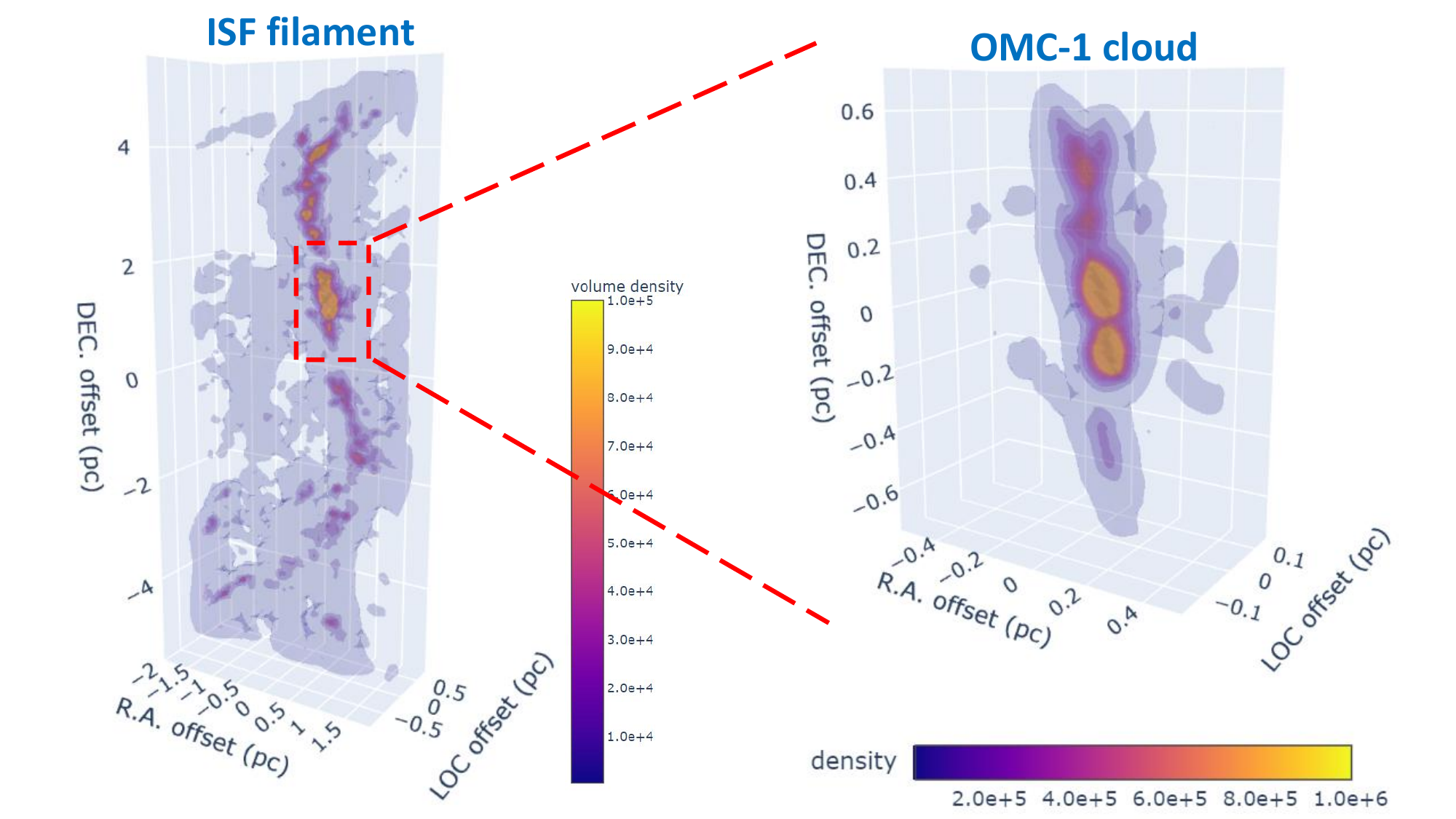}
    \caption{{\bf 3D rendering of a density reconstruction performed toward the
    Orion A molecular cloud. }The column density map is derived from Herschel observations \citep{2013ApJ...763...55R,2013ApJ...777L..33P}.}
    \label{figISF}
\end{figure}

\section{Performance \& Applications}

\begin{figure}
    \centering
    \includegraphics[width=\linewidth]{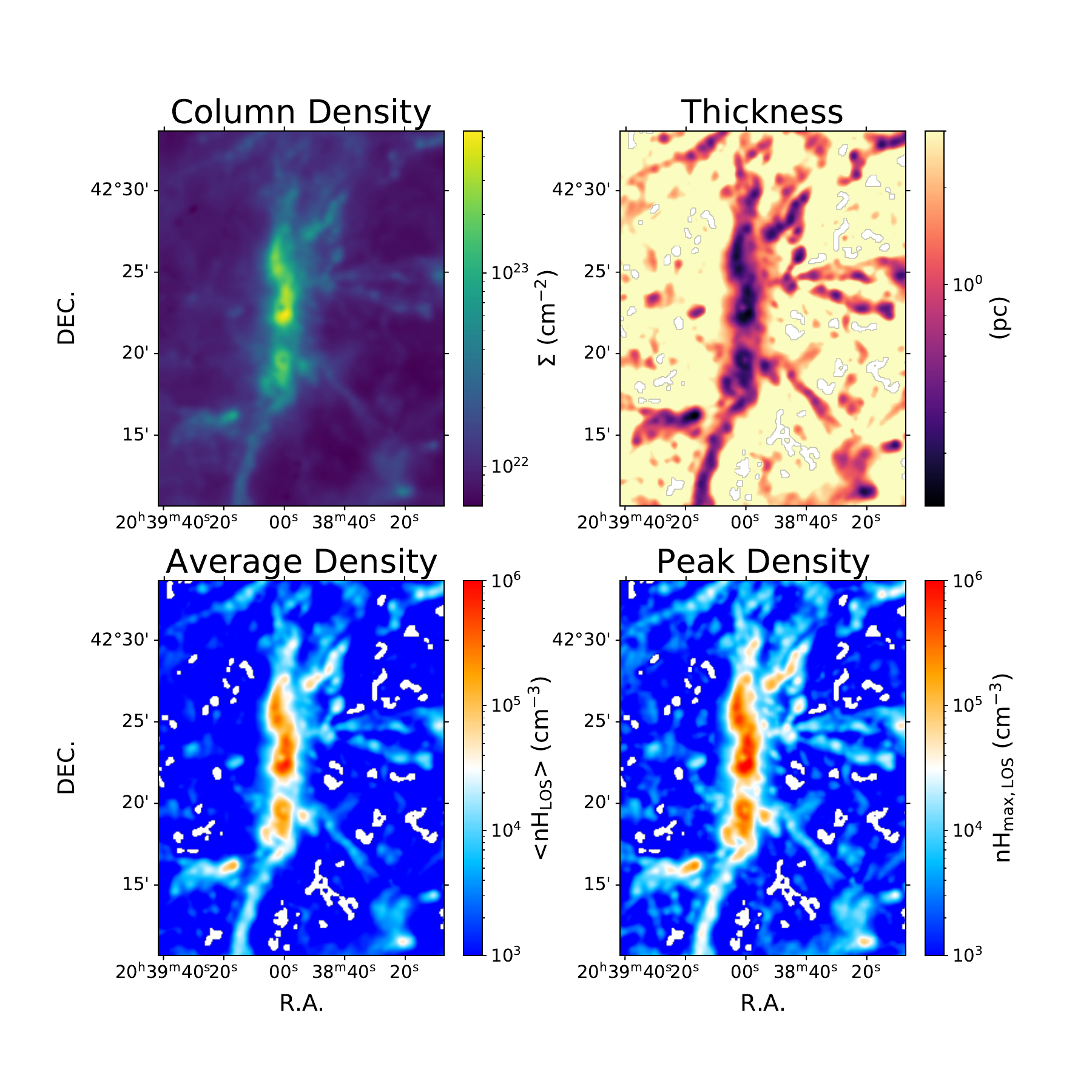}
    \caption{{\bf Information derived from 3D density reconstruction: mean density, peak density, and the thickness distribution in POS plane.}}
    \label{figDR21information}
\end{figure}


\subsection{Validation}
To evaluate the accuracy of our result, we derive the mass-weighted average
volume density $\langle\rho\rangle = \sum\rho^2\Delta V / \sum\rho\Delta V$,
the maximum density ($\rho_{\rm max}$), and the mass-weighted dispersion the
densities long different line of sights. We derive those quantities, both from 
the 3D density reconstruction and from the true 3D simulation data, and compare
the two. 

\begin{figure}
    \centering
    \includegraphics[width=0.95\linewidth]{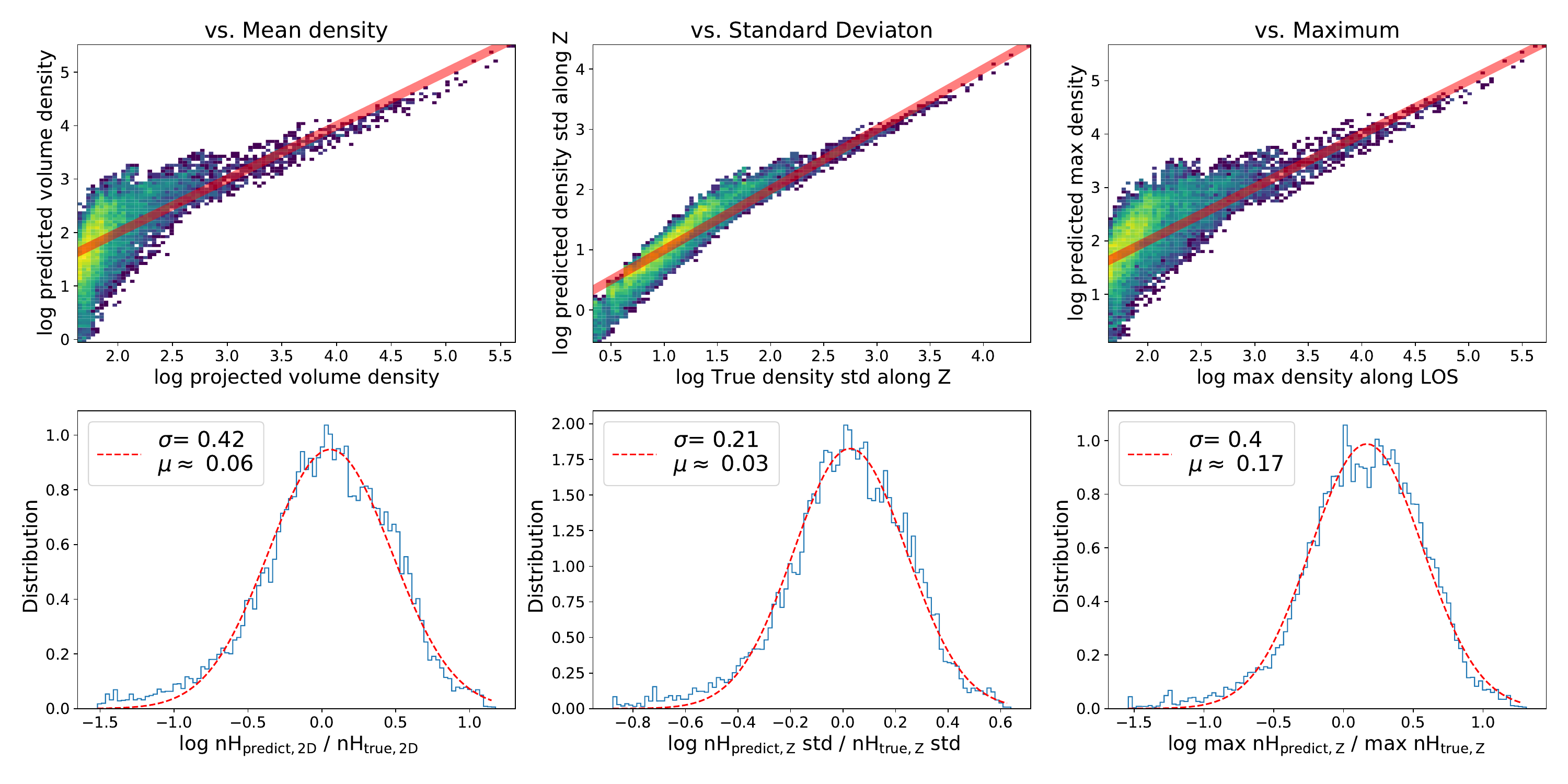}
    \caption{{\bf Validation of the predicted mean density, standard deviation, and maximum along the LOC (z axis) with true value in simulation M3.}
    The top panels display the 2D histogram between predicted and true value in simulation M3, which are mean density, standard deviation of density and maximum density along the LOC from left to right, respectively.
    The red lines shows the equality between predicted and true value.
    The bottom panels show the 1D histogram of the ratio between predicted and true value, which come from the top panels.
    the dot red lines present the gaussian fitting line to obtain the average value $\mu$ and dispersion $\sigma$.
    }
    \label{figM3z}
\end{figure}

The simulation data
\citep{2017ApJ...850...62I,2019A&A...630A..97C,2020ApJ...905...14B} we use was
produced using the \texttt{FLASH} code \citep{Fryxell2000}
. We choose a
few clouds from simulations of different evolutionary stages and magnetization,
such that our sample cover a few possible ways gas is organized:  M3
cloud mainly exhibits filamentary structures, the
M4 cloud has an extend shape with multiple sub-structures, like filaments and
clumps, M8 cloud includes high magnetic energy to simulate the strong magnetic
state of ISM. The comparison results can be seen in Fig.\,\ref{figM3z},
\ref{figM4z} and \ref{figM8z}. From these exercises, we evaluated the accuracy of
the method, which can be summarized as follows:

\begin{itemize}
\item Predicted mean volume density is close to true values in M3, M4, and M8
simulations, with a slight underestimation of around 0.1 dex, and a dispersion
of 0.3 dex.
\item Predicted density standard deviation aligns well with true values, with the ratio forming a Gaussian distribution peaking near unity with a 0.2 dex dispersion.
\item Predicted maximum density along LOS closely matches true values, with the ratio following a Gaussian distribution peaking near unity with a 0.3 dex dispersion.

\end{itemize}
The reconstruction accuracy remains consistent across different projected
directions, and the method is effective for varied structures (filamentary in M3,
complex in M4, high magnetic energy in M8).

\subsection{Rebuilding the volume density PDF}

\begin{figure}
    \centering
    \includegraphics[width=\linewidth]{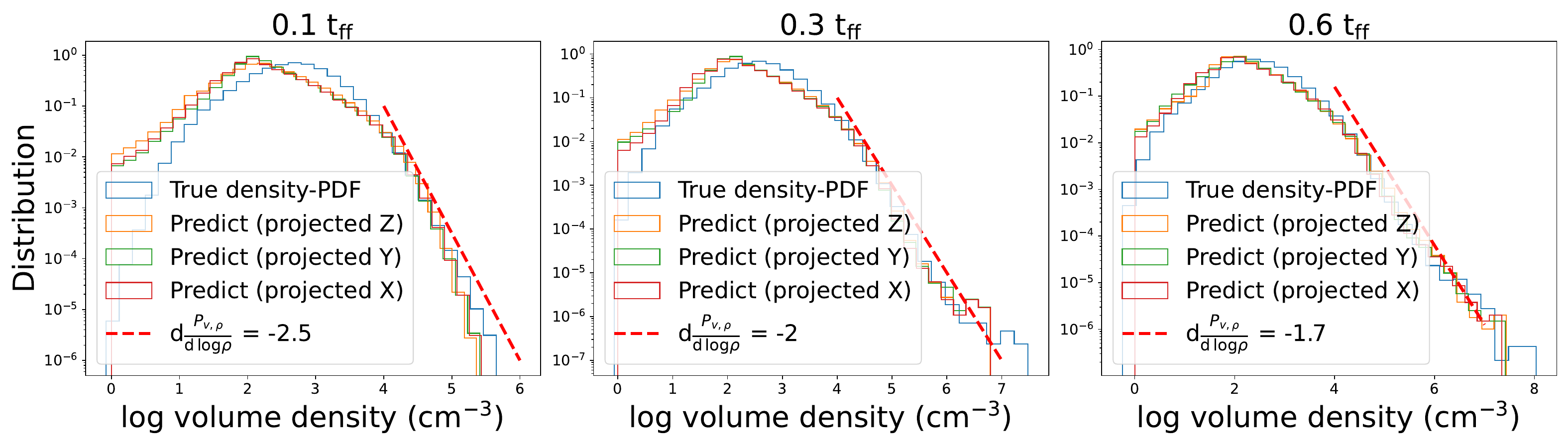}
    \caption{{Rebuilding the 3D density probability distribution function (PDF) in ENZO simulation.}
    The blue lines present the density PDF from 3D simulation.
    The panels from left to right present simulations evolved at different timescales of 0.1, 0.3, 0.6\,t$_{\rm ff}$.
    The orange, green red lines display the rebuilding density PDF from column density at different projection, along Z, Y, X axis.
    The red dot lines show the power-law tail and it slope.}
    \label{figenZoPDF}
\end{figure}

The density probability distribution function (PDF) \cite{kainulainen2009probing} is a fundamental diagnostic
of molecular cloud structure, and with the density being a key parameter of the
fluid which sets the free-fall time, the volume-density PDF is a physically
meaningful quantity whose shaped can be correlated with other measures such as
star formation to understand cloud evolution \cite{2014Sci...344..183K}. 
Our constrained-diffusion-based approach offers a new, fast way to
derive the density PDF.


We validate this approach using an \texttt{ENZO} \citep{bryan2014enzo} simulation taken at
\url{https://www.mhdturbulence.com/} \citep{Burkhart2020} performed by \citet{Collins2012,Burkhart2015}. By choosing
different snapshots ($\beta_0 = 20$, $t = 0.1, 0.3, 0.6 t_{\rm ff}$), we
evaluate if our density reconstruction can capture the evolution of the density
PDF under gravity for clouds with complex density structures that resemble
reality \citep{2024ApJ...976..209Z}. The consistency is further scrutinized by
choosing different projections. 
As Fig.\,\ref{figenZoPDF} shows, the 3D density reconstruction method can
accurately recover the log-norm distribution and the power-law tail of PDF. The
transition to power-law tail can also be captured.



\subsection{Applications}
\begin{figure}
    \centering
    \includegraphics[width=\linewidth]{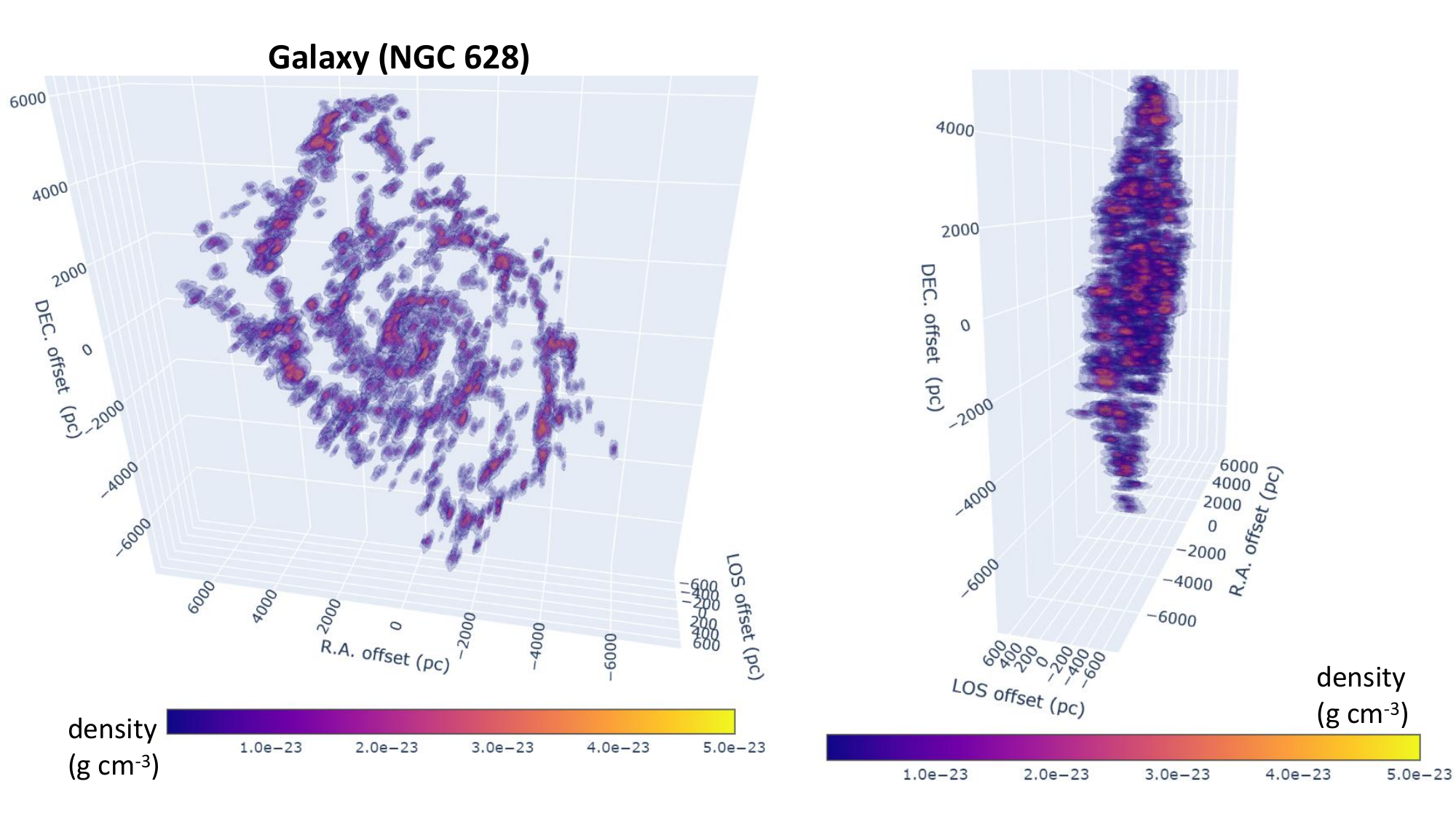}
    \caption{{\bf 3D view of kpc-scale structure, galaxy NGC 628.}
    The column density come from Phangs-ALMA survey  \citep{2021ApJS..257...43L}.}
    \label{fig:enter-label}
\end{figure}

\begin{figure}
    \centering
    \includegraphics[width=\linewidth]{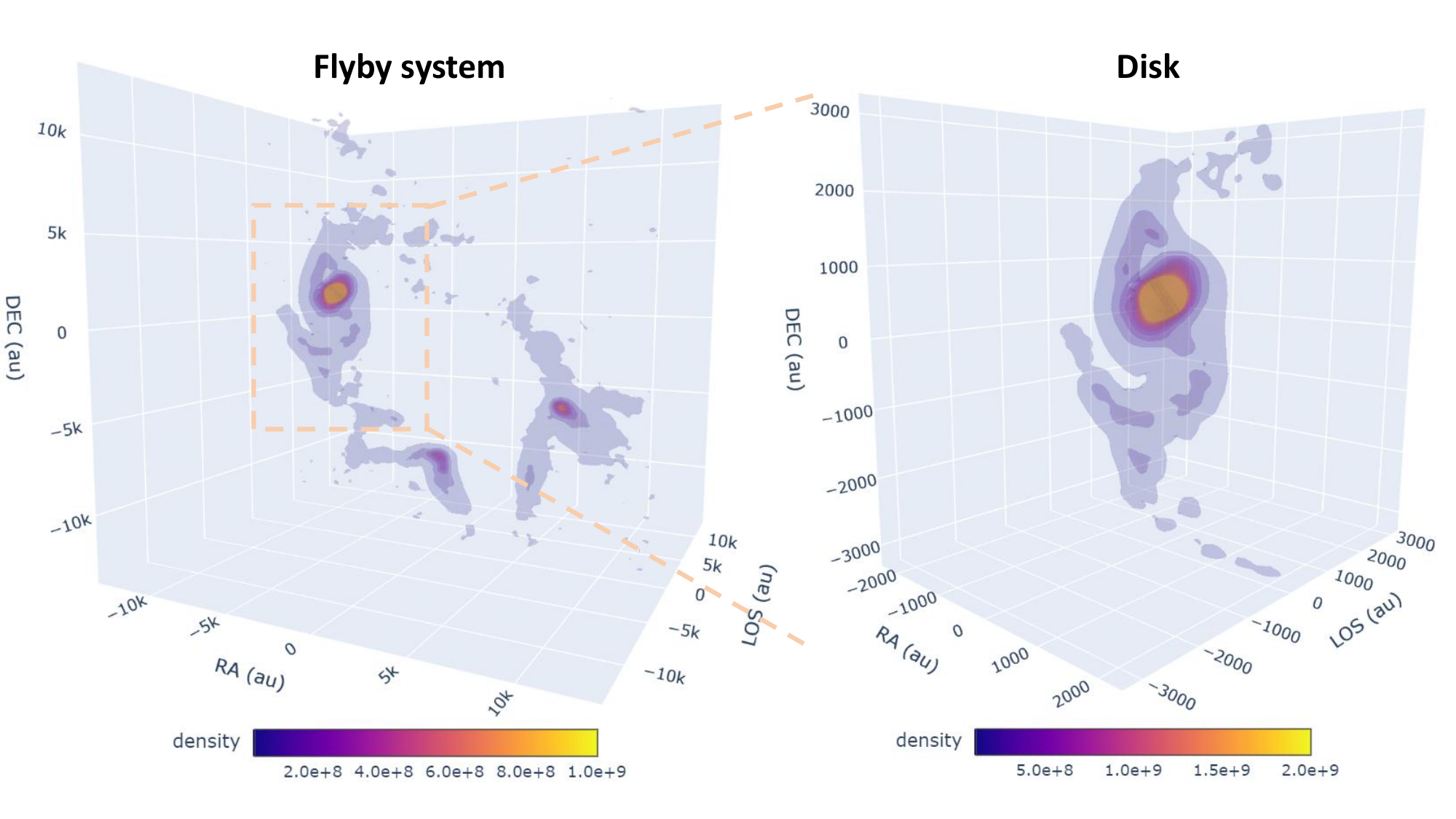}
    \caption{{\bf 3D view of kAU-scale structure, flyby system, and proto-stellar disk.}
    The raw data, column density, is from the 220\,GHz continuous  \citep{2022NatAs...6..837L}}
    \label{figflyby}
\end{figure}

To demonstrate the use of the code, we apply it to a few systems of, including a galaxy (NGC\,628, see
Fig.\,\ref{fig:enter-label}), a molecular cloud (Orion\,A, see
Fig.\,\ref{figISF}), a  proto-stellar disk (flyby system in
Sgr C, see Fig.\,\ref{figflyby}). These data are taken from the PHANGS-ALMA
survey of the nearby spiral galaxy NGC 628
\citep{2021ApJS..257...43L,2021ApJS..255...19R}, the Orion A molecular cloud
\citep{2013ApJ...777L..33P,2013ApJ...763...55R}, and the 220,GHz continuum
emission from the proto-stellar disk and flyby system
\citep{2022NatAs...6..837L}.

\section{Conclusion}

The Volume Density Mapper is a useful tool for reconstructing 3D density
distributions of the interstellar medium from 2D column density maps using
constrained diffusion. It accurately captures mean density, maximum density, and
density standard deviation, validated against FLASH and ENZO simulations. The
method applies to diverse systems like NGC 628, Orion A, and Sgr C, offering
insights into cloud structure and star formation. The performance of the method
can be summarized as follows:
\begin{itemize}
    \item Along each line of sight, the mean volume density estimates within 0.1 dex of true values, with 0.3 dex dispersion across M3, M4, and M8 simulations.
    \item Along each line of sight, the density standard deviation ratios peak near unity, with 0.2 dex dispersion.
    \item Along each line of sight, the maximum density ratios match true values, with 0.3 dex dispersion.
    \item Reconstructs 3D density PDFs, capturing log-normal and power-law components in ENZO simulations.
\end{itemize}

The code is available at \url{https://github.com/gxli/volume-density-mapper}. It outperforms wavelet-based methods for high-contrast densities distributions and is more efficient than filament-based approaches.

\bibliographystyle{aasjournal}
\bibliography{reference}

\appendix

\section{Data}

\subsection{Flash Simulation}

M3 and M4 clouds are three-dimensional numerical MHD simulations  \citep{2017ApJ...850...62I,2019A&A...630A..97C,2020ApJ...905...14B} with self-gravitating, magnetized, SN-driven, turbulent, multiphase-ISM using the using the FLASH v4.2.2 adaptive mesh refinement code (AMR;  \citealt{2000ApJS..131..273F}). 
The evolving time is around 3 Myrs and the spatial resolution is 0.06 pc.
The spatial resolution of M8 is 0.12 pc.
The M3 cloud has a classical filament structure, the M4 cloud with extend structure have multiple structure like filament and clumps, and the M8 cloud presents a strong magnetic state of ISM structure.

\subsection{Enzo Simulation}
For this study, we identified a molecular cloud through self-gravitating simulations using the constrained transport MHD option in Enzo (MHDCT) code  \citep{2012ApJ...750...13C,2015ApJ...808...48B,2020ApJ...905...14B}. 
The simulations aimed to analyze the effects of self-gravity and magnetic fields on supersonic turbulence in iso-thermal molecular clouds. 
We use the simulation data ($\beta_0$ = 20, t $\approx$ 0.76 Myr), which initial conditions are:
\begin{equation}
    {\cal M}_s = \frac{v_{\rm rms}}{c_s} = 9
\end{equation}
\begin{equation}
    \alpha_{vir} = \frac{5 v_{\rm rms}^2 }{ 3 G \rho_0 L_0^2} = 1
\end{equation}
\begin{equation}\label{eq3}
    \beta_0 = \frac{8\pi c_s^2 \rho_0}{B_0^2} = 20
\end{equation}
This simulation setup is chosen such that the results resemble the structure of an actual star-forming molecular cloud, which is super-Alfvenic and close to the high-density molecular cloud  \citep{2024ApJ...976..209Z}.

\section{Projected in other axis of M3, M4, M8}

\begin{figure}
    \centering
    \includegraphics[width=0.95\linewidth]{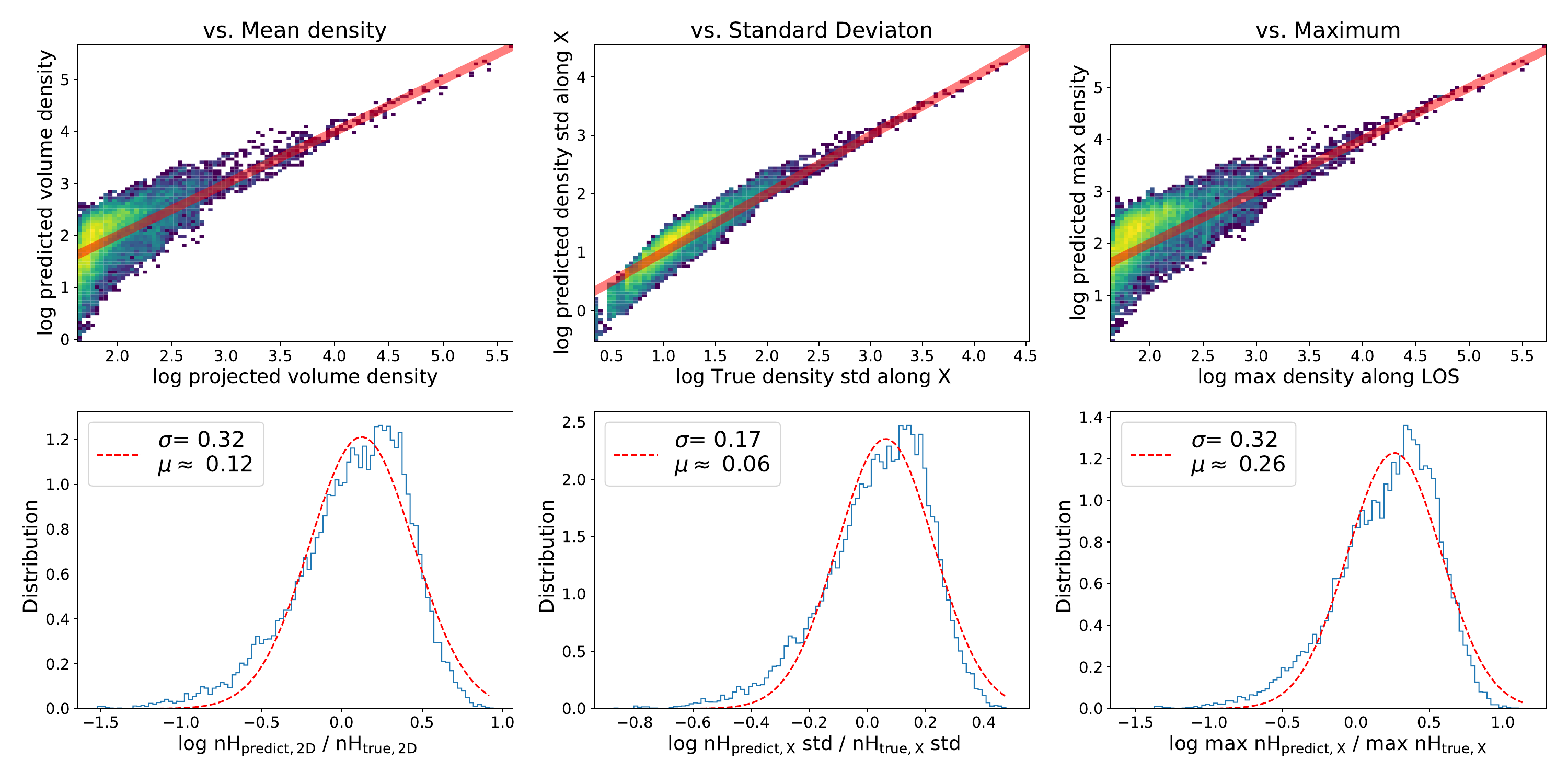}
    \includegraphics[width=0.95\linewidth]{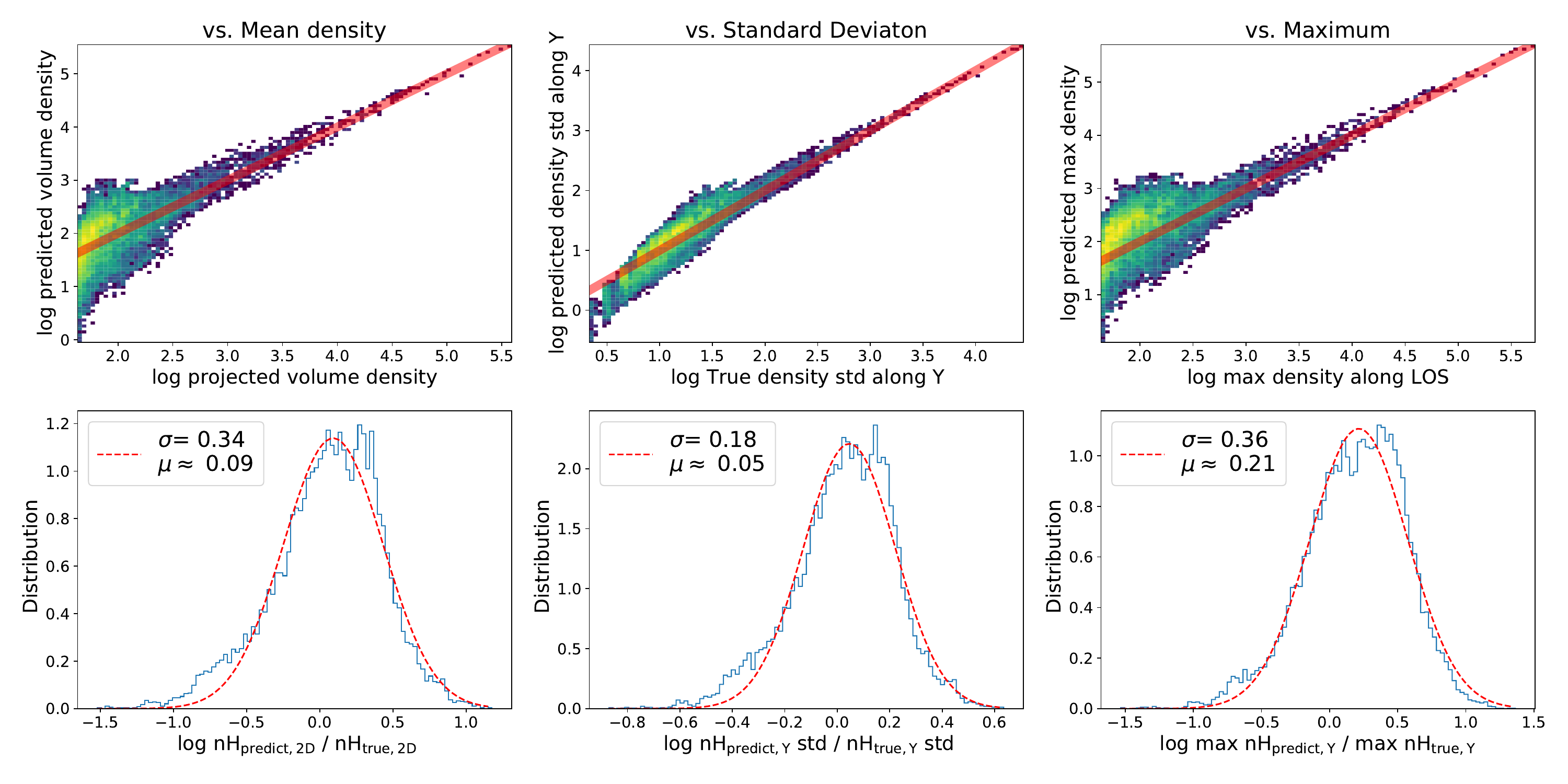}
    \caption{{\bf Validation of the predicted mean density, standard deviation, and maximum along the LOC (x,y axis) with true value in simulation M3.}}
    \label{figM3xy}
\end{figure}

\begin{figure}
    \centering
    \includegraphics[width=0.95\linewidth]{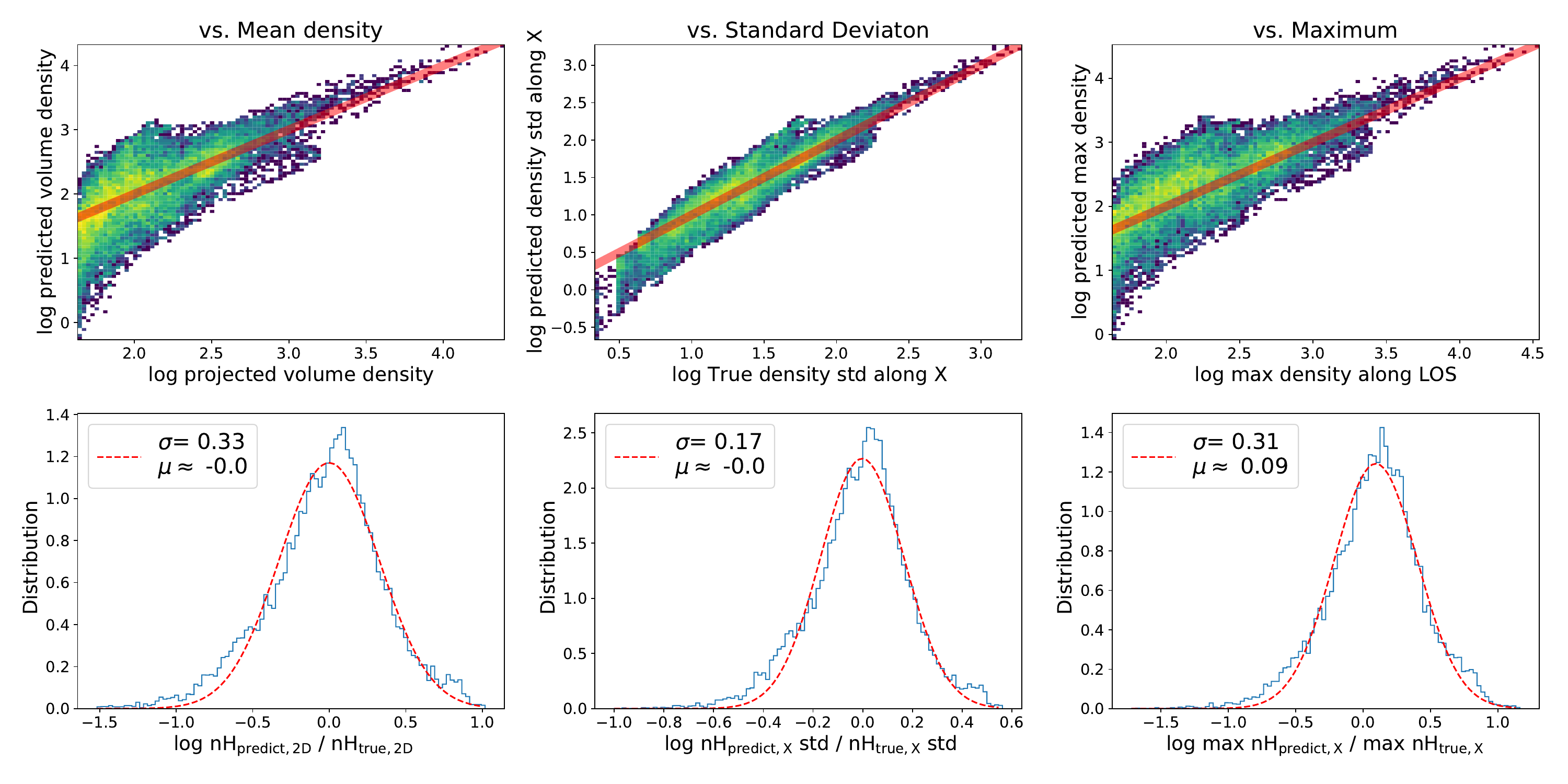}
    \includegraphics[width=0.95\linewidth]{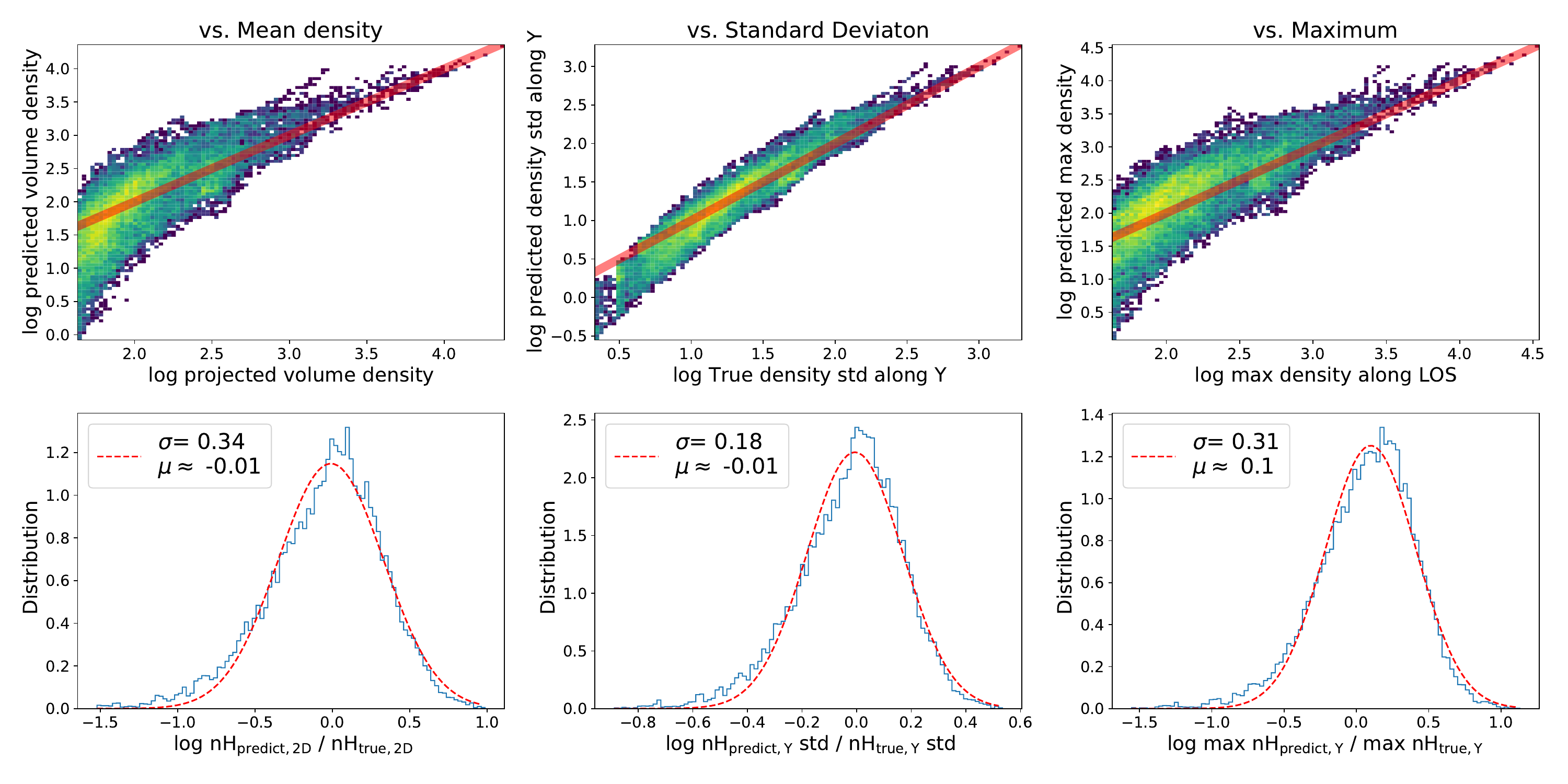}
    \caption{{\bf Validation of the predicted mean density, standard deviation, and maximum along the LOC (x,y axis) with true value in simulation M4.}}
    \label{figM4xy}
\end{figure}

\begin{figure}
    \centering
    \includegraphics[width=0.95\linewidth]{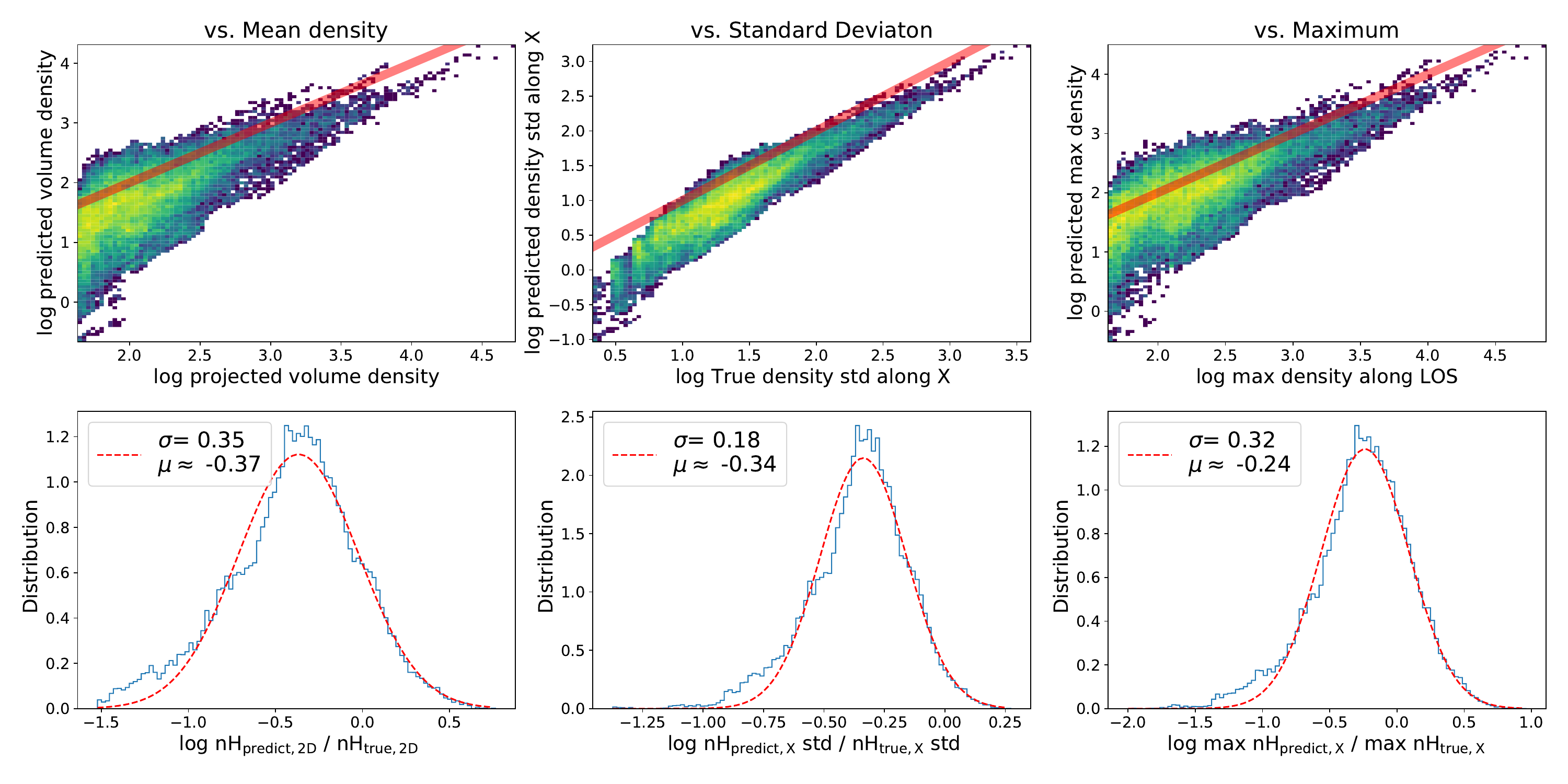}
    \includegraphics[width=0.95\linewidth]{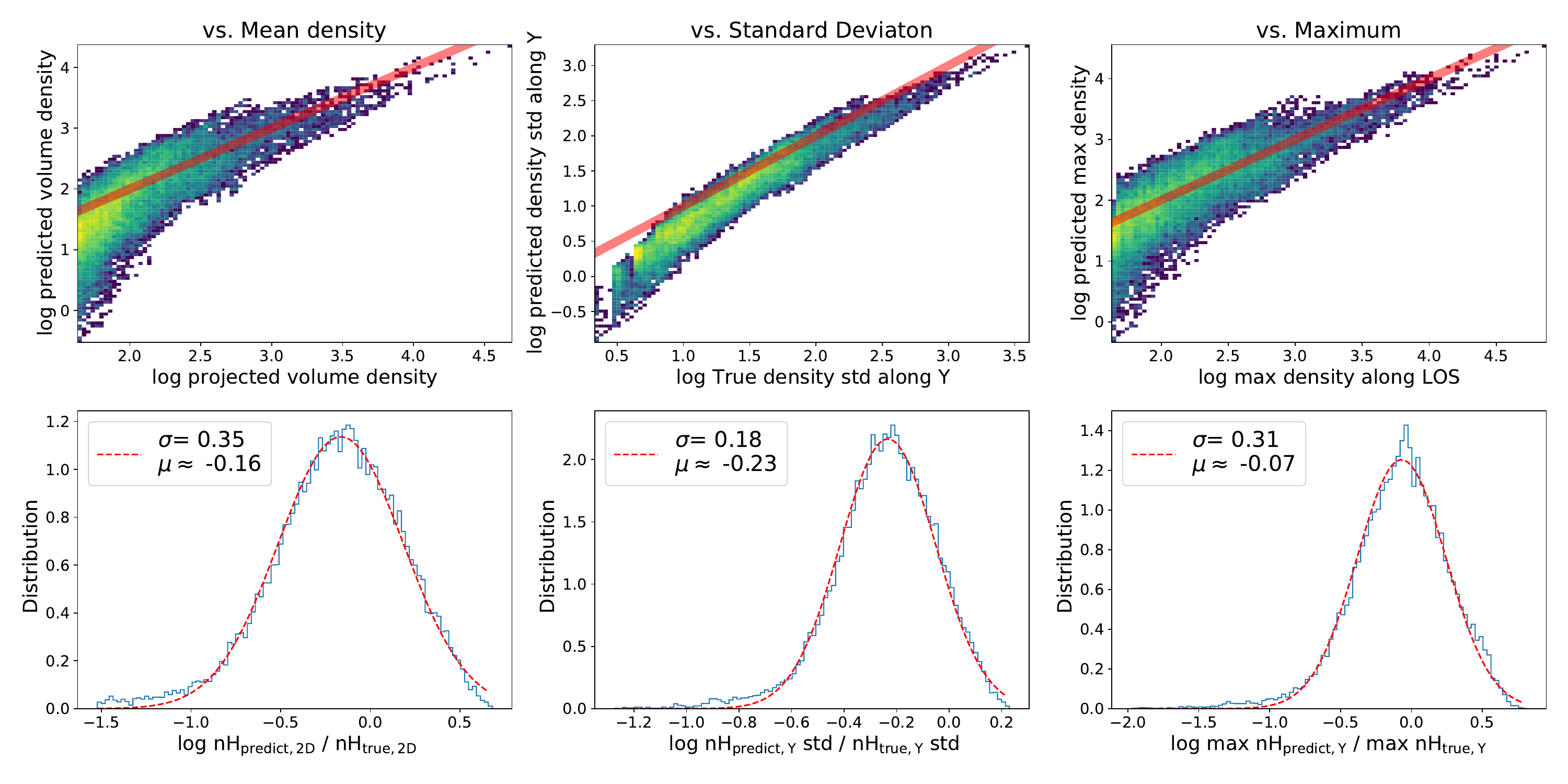}
    \caption{{\bf Validation of the predicted mean density, standard deviation, and maximum along the LOC (x,y axis) with true value in simulation M8.}}
    \label{figM8xy}
\end{figure}

This section shows the other projected direction along the x and y-axis in cloud M3, M4, and M8.
The predicted volume density also agree to the true volume density in 2D plane (see Fig.\,\ref{figM3xy}, \ref{figM4xy} \ref{figM8xy}), which does not depended on the special projected direction.

\begin{figure}
    \centering
    \includegraphics[width=0.95\linewidth]{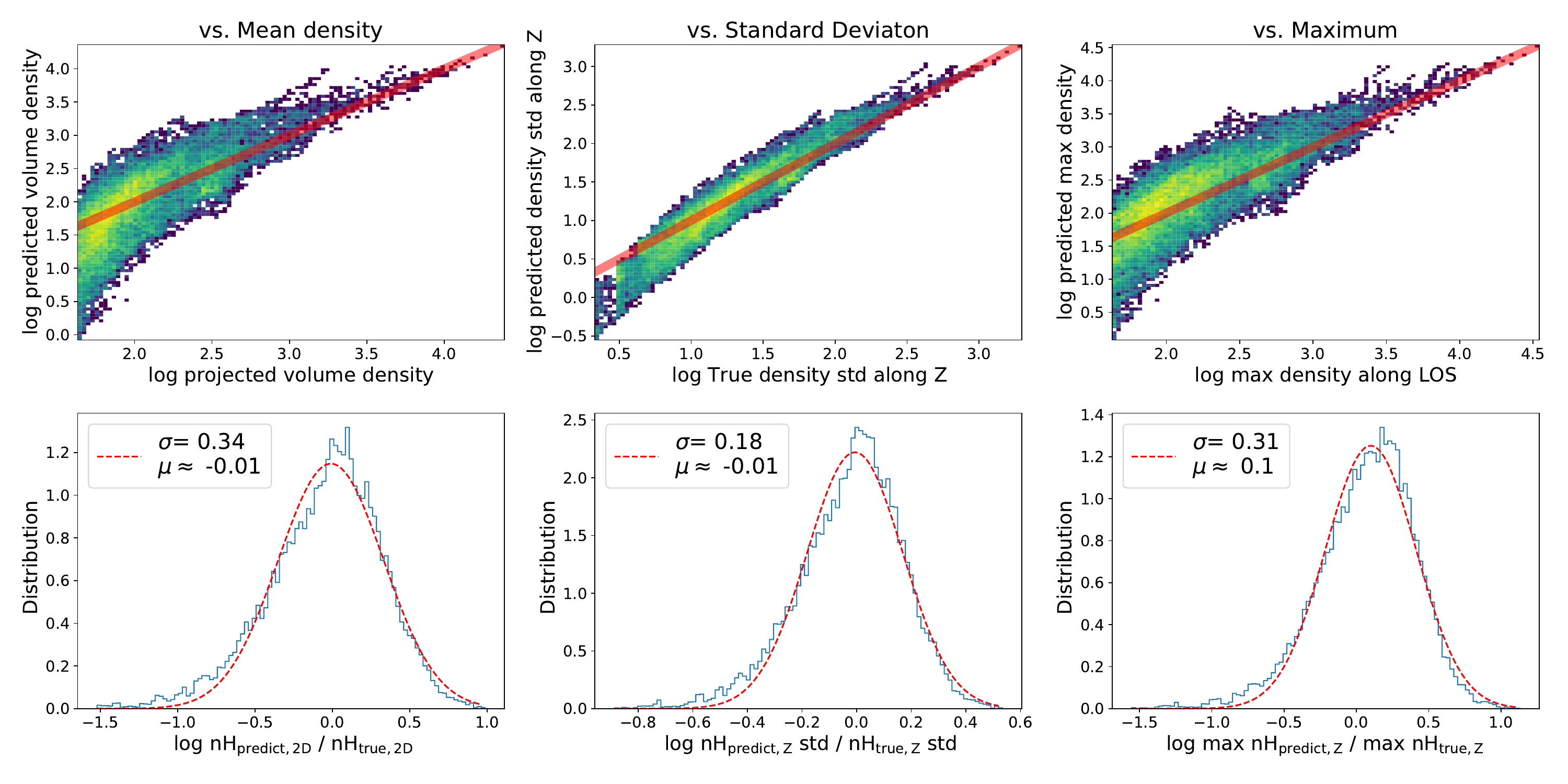}
    \caption{{\bf Validation of the predicted mean density, standard deviation, and maximum along the LOC (z axis) with true value in simulation M4.}
    These panels is the same as Fig.\,\ref{figM3z} but for M4 cloud.
    }
    \label{figM4z}
\end{figure}

\begin{figure}
    \centering
    \includegraphics[width=0.95\linewidth]{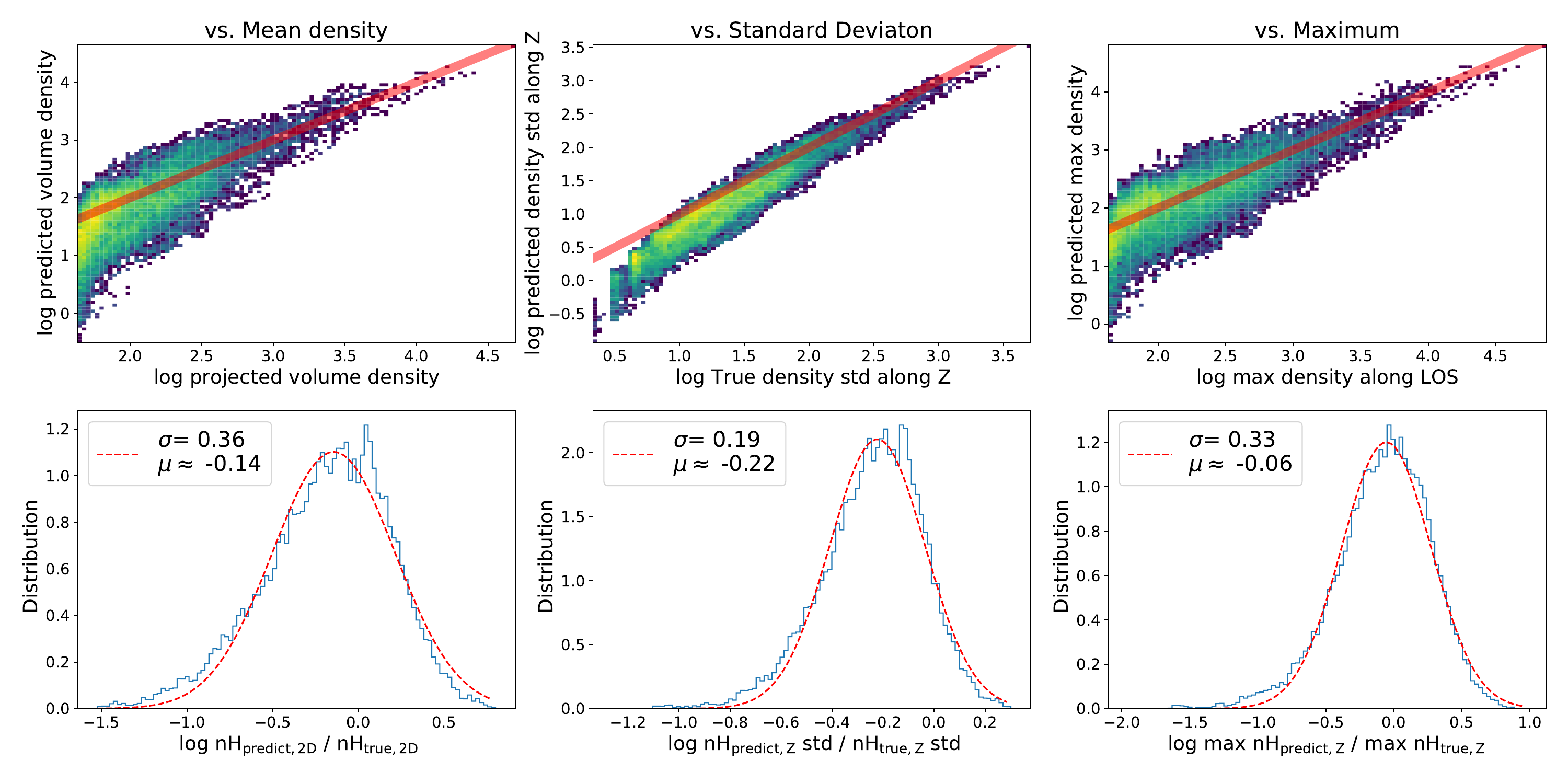}
    \caption{{\bf Validation of the predicted mean density, standard deviation, and maximum along the LOC (z axis) with true value in simulation M8.}
    These panels is the same as Fig.\,\ref{figM3z} but for M8 cloud.
    }
    \label{figM8z}
\end{figure}

\end{document}